\documentclass[english,aps,prb,twocolumn]{revtex4}

\usepackage{amssymb}
\usepackage{color}
\usepackage{array}
\usepackage{graphicx}
\usepackage{babel}

\usepackage{hyperref}
\hypersetup{colorlinks=false}

\begin{document}

\title{Disorder-induced cavities, resonances, and lasing in randomly-layered
media}
\author{Yury Bliokh}
\affiliation{Department of Physics, Technion-Israel Institute of Technology, Haifa 32000,
Israel,}
\affiliation{Advanced Science Institute, RIKEN, Wako-shi, Saitama 351-0198, Japan}
\author{Elena I. Chaikina}
\affiliation{Divisi\'on de F\' \i sica Aplicada, Centro de Investigaci\'on Cient\' \i fica y de Educaci\'on Superior de Ensenada. Carretera Ensenada-Tijuana No. 3918, Ensenada, BC, 22860 M\'exico.}
\author{Noem\'i Liz\'arraga}
\affiliation{Divisi\'on de F\' \i sica Aplicada, Centro de Investigaci\'on Cient\' \i fica y de Educaci\'on Superior de Ensenada. Carretera Ensenada-Tijuana No. 3918, Ensenada, BC, 22860 M\'exico.}
\author{Valentin Freilikher}
\affiliation{Department of Physics, Jack and Pearl Resnick Institute, Bar-Ilan
University, Israel}
\affiliation{Advanced Science Institute, RIKEN, Wako-shi, Saitama 351-0198, Japan}
\author{Eugenio R. M\'endez}
\affiliation{Divisi\'on de F\' \i sica Aplicada, Centro de Investigaci\'on Cient\' \i fica y de Educaci\'on Superior de Ensenada. Carretera Ensenada-Tijuana No. 3918, Ensenada, BC, 22860 M\'exico.}
\author{Franco Nori}
\affiliation{Advanced Science Institute, RIKEN, Wako-shi, Saitama 351-0198, Japan}
\affiliation{Department of Physics, University of Michigan, Ann Arbor, Michigan
48109-1040, USA }

\begin{abstract}
We study, theoretically and experimentally, disorder-induced resonances in
randomly-layered samples,and develop an algorithm for the detection and
characterization of the effective cavities
that give rise to these resonances. This algorithm enables us 
to find the eigen-frequencies and pinpoint the locations of the resonant
cavities that appear in individual realizations of random samples,
for arbitrary distributions of the widths and refractive indices of the layers. Each cavity is formed in a region whose size is a few
localization lengths. Its eigen-frequency is independent of the location
inside the sample, and does not change if the total length of the sample is
increased by, for example, adding more scatterers on the sides. We show that
the total number of cavities, $N_{\mathrm{cav}}$, and resonances, $N_{\mathrm{res}}$, per unit
frequency interval is uniquely determined by the size of the disordered
system and is independent of the strength of the disorder. In an active,
amplifying medium, part of the cavities may host lasing modes whose number
is less than $N_{\mathrm{res}}$. The ensemble of lasing cavities behaves as distributed feedback lasers, provided that the gain of the medium exceeds the lasing threshold, which is specific for each cavity.
We present the results of experiments carried out with 
single-mode optical fibers with gain and randomly-located resonant Bragg
reflectors (periodic gratings). When the fiber was illuminated by a pumping
laser with an intensity high enough to overcome the lasing threshold, the
resonances revealed themselves by peaks in the emission spectrum. Our experimental results are in a good agreement with the theory presented here.
\end{abstract}

\maketitle

\section{Introduction}

Anderson localization of waves in one-dimensional random media is associated
with an exponential decrease of the wave amplitude inside the disordered,
locally-transparent sample, which results in an exponentially small
typical transmittance {$T_{\mathrm{typ}}\varpropto \exp \left( -L/l_{\mathrm{loc}}\right) \ll 1$} (where $L$ is the length of the sample, and {$l_{\mathrm{loc}}$} is the localization length). Another manifestation of Anderson
localization is the existence of resonant frequencies, where the
transmission increases drastically, sometimes up to unity. These frequencies
correspond to the quasi-localized eigenstates (modes, or disorder-induced
resonances) characterized by a high concentration of energy in
randomly-located points inside the system.

Even though one-dimensional (1D) localization has been intensively studied
during the last few decades \cite{50 years} (see also \cite{LGP, Ping} and
references therein), most of the analytical results were obtained for mean
quantities, i.e., for values averaged over ensembles of random realizations.
These results are physically meaningful for the self-averaging Lyapunov
exponent (inverse localization length), which becomes non-random in the
macroscopic limit \cite{LGP}. For non-self-averaging quantities (field
amplitude and phase, intensity, transmission and refection coefficients,
etc.) a random system of any size is always mesoscopic, and therefore, mean
values have little to do with measurements at individual samples. This
is most pronounced when it comes to the disorder-induced resonances whose
parameters are extremely volatile and strongly fluctuate from realization to
realization \cite{Azbel1,Azbel2}. In particular, ensemble averaging wipes
out all information about the frequencies and locations of individual
localized states within a particular sample, even though these data are
essential for applications based on harnessing micro and nano cavities with
high $Q$-factors.

\subsection{Disorder-induced cavities and resonances}

Nowadays, photonic crystals are believed to be the most suitable platforms
for the creation and integration of optical resonators into optical
networks. To create an effective resonant cavity that supports localized
high-$Q$ modes in a photonic crystal (PC), it is necessary to break
periodicity, i.e. to introduce a defect in a regular system. As fluctuations
of the dielectric and geometrical parameters are inevitably present in any
manufactured periodic sample, it could create a serious obstacle in the
efficient practical use of PCs. Therefore considerable efforts of researches
and manufacturers go into the control of fluctuations. Alternatively, if
rather than combating imperfections of periodicity, one fabricated
highly-disordered samples, they could be equally well harnessed, for
example, for creating tunable resonant elements. This is because 1D random
configurations have a unique band structure that for some applications have
obvious advantages over those of PC \cite{Genack we 1}. Contrary to periodic
systems, resonant cavities inherently exist in any (long enough) disordered
sample, or can be easily created by introducing a non-random element (for
example, homogeneous segment) into an otherwise random configuration. The
effective wave parameters of these cavities are very sensitive to the fine
structure of each sample, and can be easily tuned either by slightly varying
the refractive index in a small area inside the sample or, for example, by
changing the ratio between the coupling strength and absorption. This makes
possible to shift the resonant frequency (and thereby to lock and unlock the
flow of radiation); to couple modes and create quasi-extended states, etc 
\cite{Genack we 2, Labonte}.

It has been shown \cite{JOSA} that each localized state at a frequency $%
\omega =\omega _{\mathrm{res}}$ could be associated with an effective, high
$Q$-factor resonance cavity comprised of almost transparent (for this
frequency) segment bounded by essentially non-transparent regions (effective
barriers). Wave tunneling through such a system can be treated as a
particular case of the general problem of the transmission through an open
resonator \cite{Colloquium}. The distinguishing feature of a
disorder-induced cavity is that it has no regular walls (the medium is
locally transparent in each point), and high reflectivity of the confining
barriers is due to the Anderson localization. Moreover, different segments
of the sample are transparent for different frequencies, i.e., each
localized mode is associated with its own resonator.

\subsection{Random lasers}

If the medium inside such a cavity is amplifying, the combination of the
optical gain and the interference of multiply scattered radiation creates
coherent field and gives rise to multi-frequency lasing with sharply peaked
lasing spectrum. Random lasers (RL) are the subject of increasing
scientific interest due to their unusual properties and promising potential
applications \cite{Cao1, Noginov, Cao99, Ara, Patrick new 1, Patrick new 2,
RLasers}. Unlike conventional lasers, where any disorder is detrimental, in
a random version, scattering plays a positive role increasing the path
length and the dwell time of light in the active medium. So far most
studies, both experimental and theoretical, have concentrated on 3D
disordered systems and chaotic cavities. One of the grave drawbacks of 3D
random lasers is their inefficient pumping, which is hampered by the
scattering of the pumping radiation in the random medium. A one-dimensional
RL, which is free from this disadvantage, can be realized either as a random
stack of amplifying layers \cite{Genack2004} or as a set of Bragg gratings
randomly distributed along a doped optical fiber \cite{Noemi}. In the last
case, the wavelength of the pumping laser is shifted from the Bragg
resonance of the gratings, and the fiber is excited homogeneously along the
whole length of the sample. Both these methods reduce noticeably the lasing
threshold as compared to 3D random lasing systems. Because the frequencies
of the modes and the locations of the effective cavities vary from sample to
sample randomly, in the most cases they are described statistically \cite%
{Misirpashaev98,Zaitsev06,Zaitzev09,Soukoulis2000,Cao2007}. However, usually
we deal with a specific random sample, and it is important to know how many
modes and at which frequencies can be excited in a given frequency range;
where these modes are localized inside this sample, etc \cite{Patrick
new 1}.

Another field of research in which this information is crucial has arisen
recently after it had been realized that Anderson resonances could be used
to observe cavity quantum electrodynamics effects by embedding a single
quantized emitter (quantum dot) in a disordered PC waveguide \cite%
{Lodahl-1,Lodahl-2}. In those experiments, the efficiency of the interaction
between radiation and disorder-induced cavities depends strongly on the
location of the source inside the random sample. Indeed, all QED effects are
well-pronounced when the emitter with a given frequency is placed inside the
effective cavity, which is resonant at this frequency, and could be
completely suppressed otherwise.

In this paper, we develop an algorithm that enables to detect all cavities
and to find their locations and eigen-frequencies for any \textit{individual}
sample with given geometry and optical parameters. It is shown that in the
case of uncorrelated disorder, the number of disorder-induced resonances per
unit frequency interval is independent of the strength of the fluctuations
and is uniquely determined by the size of the random sample. The results
have been checked experimentally using RLs based on a single-mode fiber with
randomly distributed resonant Bragg gratings developed in \cite{Noemi}.

\section{Disorder-induced cavities and resonances \label{Definition}}

It has been shown in \cite{JOSA,Colloquium,Genack we 1,Genack we 2} that for
a quantitative description of the wave propagation through a disordered
sample, it is advantageous to consider it as a random chain of effective
regular resonators with given Q-factors and coupling coefficients. The
typical size of each resonator is of the order of the localization length, $%
l_{\mathrm{loc}}$, and their centers are randomly distributed along the
sample. For manufacturing RLs and for the ability to tailor their
properties, it is important to know the location of the resonant cavity for
each eigen-frequency \cite{Patrick new 1}. To this end, a criterion is
necessary, which enables to determine whether a given area of a disordered
sample is either a resonant cavity or a strongly reflecting (typical) random
segment.

To derive such a criterion for randomly layered media, we consider a
disordered sample consisting of $N$ \ homogeneous layers with thicknesses $%
d_{j}$ and dispersionless refractive indices $n_{j}$ ($j=1,2,\ldots ,N$)
that are statistically independent and uniformly distributed in the
intervals $(d_{0}-\delta \!d,d_{0}+\delta \!d)$ and ($n_{0}-\delta
\!n,n_{0}+\delta \!n)$, respectively. (The effects of correlation between
the thicknesses of the adjacent blocks have been considered in \cite{FN 1,FN
2,FN 3}). In a system with uncorrelated layers, the interface between the $j$%
-th and $(j+1)$-th layers is located at a random point $z_{j}$ and can be
characterized by complex transmission, $t_{j}$, and reflection, $r_{j}$,
coefficients, which are also randomly distributed in the corresponding
intervals. The numeration is chosen from left to right so that $%
z_{j}<z_{j+1} $. It is assumed thereafter that the optical contrast between
layers is small. Therefore all Fresnel reflection coefficients $r_{j} $ are
also small 
\begin{equation}
|r_{j}|\ll 1  \label{small r}
\end{equation}%
and, consequently, $1-|t_{j}|^{2}\ll 1$. For the sake of simplicity these
coefficients will be assumed real. Generalization to complex-valued $r_{j}$
and $t_{j}$ is straightforward.

In what follows, it is convenient to turn to\ the \textit{optical lengths},
i.e. to replace the thicknesses $d_{j}$ and the coordinates $z_{j}$ by 
\begin{eqnarray}
\tilde{d}_{j} &=&n_{j}d_{j},  \nonumber  \label{opt d} \\
\tilde{z}_{j} &=&\sum_{i<j}n_{i}z_{i},
\end{eqnarray}%
where $n_{j}$ is the refraction index of the $j$-th layer. Note that in
these notations the wave number has the same value $\widetilde{k}=k=\omega/c$
in all layers.

\begin{figure}[htb]
\centering \scalebox{0.3}{\includegraphics{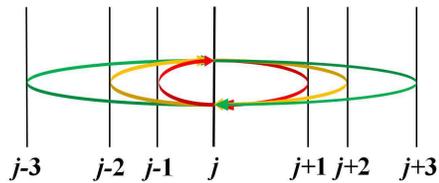}}
\caption{(color online) Schematic diagram of closed trajectories in a randomly-layered
sample.}
\label{Fig_1}
\end{figure}

To distinguish between an effective resonator and a Bragg reflector, a
proper physical quantity representative of distinctive properties of these
objects has to be found. To do this, let us consider a wave that propagates
rightward from a point $\tilde{z}=\tilde{z}_{j}+0$ [right side of the
interface between the $(j-1)$-th and $j$-th layers], where its amplitude is $%
A_{j0}^{(+)}$ (thereafter superscripts $(+)$ and $(-)$ denote the amplitudes
of the waves propagating to the right or to the left, respectively). After
traveling through the $j$-th layer, the wave is partially reflected back
from the interface between the $j$-th and $(j+1)$-th slabs. The transmitted
part extends through $(j+1)$-th layer and is partially reflected from the
next interface, etc. (see Fig.~\ref{Fig_1}).

If the scattering at the interfaces between layers is weak, Eq.~(\ref{small
r}), the amplitude $\tilde{A}_{j}^{(-)}$ of the total field reflected from $%
N_{s}$ layers and returned back to the point $\tilde{z}=\tilde{z}_{j}+0$ can
be calculated in the single-scattering approximation, and is equal to: 
\begin{equation}
\tilde{A}_{j}^{(-)}=r_{j}^{(+)}(k)\,A_{j0}^{(+)},  \label{eq1}
\end{equation}
where 
\begin{equation}
r_{j}^{(+)}(k)=\sum_{m=1}^{N_{s}}r_{j+m}\,\exp\left[2ik\left( \tilde{z}%
_{j+m}-\tilde{z}_{j}\right)\right],  \label{eq2}
\end{equation}
and $k=\omega /c$ is the wave number. The amplitude $\tilde{A}%
_{j}^{(+)}=r_{j}^{(-)}(k)\,A_{j0}^{(-)}$ is introduced in the same way, with
the \textit{left} reflection coefficient $r_{j}^{(-)}(k)$ 
\begin{equation}
r_{j}^{(-)}(k)=-\sum_{m=0}^{N_{s}}r_{j-m}\;\exp\left[2ik\left( \tilde{z}_{j}-%
\tilde{z}_{j-m}\right)\right].  \label{eq3}
\end{equation}
It is taken into account in Eq.~(\ref{eq3}) that the reflection coefficients 
$r_{l}$ and $r_{r}$ for waves that are incident on the same interface from
the left or from the right have opposite signs: $r_{l}=-r_{r}$. The field of
all waves that made a \textit{closed path} and returned back after consequent
reflections from $N_{s}$ layers located on the right, and $N_{s}$ layers
located on the left from the $j$-th layer, has an amplitude $A_{j1}^{(+)}$.
In the general case, $A_{j1}^{(+)}$ is not equal to the initial amplitude $%
A_{j0}^{(+)}$, and the difference is 
\begin{equation}
\delta A_{j}^{(+)}=A_{j0}^{(+)}-A_{j1}^{(+)}\equiv A_{j0}^{(+)}\left[
1-\Delta _{j}(k)\right] ,  \label{eq4}
\end{equation}%
where 
\begin{equation}
\Delta _{j}(k)=r_{j}^{(+)}(k)\;r_{j}^{(-)}(k).  \label{delta}
\end{equation}%
The function $\Delta _{j}(k)$ is an important characteristic, which uniquely
determines the resonant properties of any one-dimensional wave system. In
particular, the eigen-numbers can be found as poles of the Green function,
i.e. as the roots of the equation \cite{brekhovsk, Burin2002} 
\begin{equation}
\Delta _{j}(k)-1\equiv r_{j}^{(+)}(k)\,r_{j}^{(-)}(k)-1=0.  \label{Im delta}
\end{equation}

In the case of a closed resonator 
\[
\Delta (k)=\exp\left(2i\pi n\right),\hspace{3mm}n=1,2,3\ldots , 
\]
i.e. $\arg \,\Delta (k)=2\pi n$, and $\mathrm{Re}\,\Delta (k)=1$. For an
open resonant cavity, the roots $k_{\mathrm{res}}$ of  Eq.~(\ref{Im delta}%
) are complex, $k_{\mathrm{res}}=k_{R}+ik_{I}$. If the Q-factor of a
resonator is large enough, then $k_{R}\gg k_{I}$ and 
\begin{eqnarray}
\mathrm{Re}\,\Delta (k_{R}) &=&1-k_{R}\ell /Q\ >0,  \label{DeltaRe} \\
\mathrm{Im}\,\Delta (k_{R}) &\simeq &0,  \label{DeltaIm}
\end{eqnarray}%
where $\ell $ is the resonator length.

Note that equation (\ref{DeltaIm}) is fulfilled also when $\arg \Delta
(k)=\pi n$, which is the Bragg reflection condition. In this case, in
contrast to a resonant cavity, $\mathrm{sgn}\{r^{(+)}(k)\}=-\mathrm{sgn}%
\{r^{(-)}(k)\}$ and the real part of $\Delta(k)$ is negative, $\mathrm{Re}%
\,\Delta(k)<0$.

These properties of the quantity $\Delta (k)$ are quite general and can be
used to characterize randomly layered systems, in particular, to detect
effective resonant cavities inside them. Indeed, when for a segment of $%
2N_{s}$ layers centered at a point $\tilde{z}_{j}$\ inside a long ($N\gg
N_{s}$) disordered sample the imaginary part of $\Delta (k)$ is zero at some 
$k=\omega /c$, this area is either a resonant (at the frequency $\omega $)
cavity or a localization-induced resonant Bragg reflector. What it is indeed
is determined by the sign of the real part of $\Delta $: in a resonator $%
\mathrm{Re}\,\Delta >0$, while for a Bragg grating $\mathrm{Re}\,\Delta <0.$

The last condition is easy to understand if we notice that $r_{j}^{(+)}(k)$ is $(-2k)$-Fourier harmonics of the function 
\begin{equation}  \label{Distribution_1}
F_j^{(+)}(z)=\sum_{m=1}^{N_{s}}r_{j+m}\,\delta \lbrack \tilde{z}-(\tilde{z}%
_{j+m}-\tilde{z}_{j})],
\end{equation}
and $r_{j}^{(-)}(k)$ is the $(+2k)$-Fourier harmonics of the
function 
\begin{equation}  \label{Distribution_2}
F_j^{(-)}(z)=\sum_{m=0}^{N_{s}}(-r_{j-m})\,\delta \lbrack \tilde{z}-(\tilde{z%
}_{j-m}-\tilde{z}_{j})].
\end{equation}
This means that $r_{j}^{(+)}(k)$ and $r_{j}^{(-)}(k)$ are
Bragg reflection coefficients from $N_{s}$ slabs located to the right and to
the left from the $j$-th layer, respectively.

Since our prime interest here are disorder-induced resonators for which $%
\left\vert r_{j}^{(\pm )}\right\vert \sim 1$, and because in the localized
regime the reflection coefficient from a typical region is close to unity
when its length is of the order (and larger) of the localization length, for
further calculations we have to choose $N_{s}=l_{\mathrm{loc}}/d_{0}\equiv
N_{\mathrm{loc}}$. Important to stress that each resonator is formed in an
area of the size of a few localization lengths and is practically
unaffected by the outer (to this area) parts of the sample. This means that
the eigen-frequency of an effective cavity is independent of it location
inside a sample, and does not change if the total length of the sample is
increased, for example, by adding more scatterers at its edges. When the
thicknesses of the layers are uncorrelated and the mean thickness $\overline{%
\tilde{d_{j}}}=\widetilde{d}_{0}$ is large compared to the wavelength, $k%
\widetilde{d_{0}}\gg 2\pi $, the localization length is determined by the
mean value of the local reflection coefficients $r_{j}$ \cite{Lianskii}. To
prove this statement, note that the functions $r_{j}^{(\pm )}(k)$, Eqs.~(\ref%
{eq2},\ref{eq3}), are sums of $N_{\mathrm{loc}} $ uncorrelated random
complex numbers, and therefore can be described in terms of a random walk on
the plane $(\mathrm{Re}\,r,\,\mathrm{Im}\,r),$ with the single step equal to 
$|r_{j}|.$ Then, the mean absolute value $\overline{|r_{j}^{(\pm )}(k)|}$ is
the mean distance from the origin after $N_{\mathrm{loc}}$ steps: $\overline{%
|r_{j}^{(\pm )}(k)|}=\overline{|r_{j}|}\sqrt{N_{\mathrm{loc}}}\sim 1$, and
therefore 
\begin{equation}
N_{\mathrm{loc}}\simeq \left( \overline{|r_{j}|}\right) ^{-2}.  \label{eq12}
\end{equation}%
The characteristic scale of the variation of $r_{j}^{(\pm )}(k)$ along the
sample measured in number of layers is of the order of $N_{\mathrm{loc}}\gg
1 $. Thus, the coefficients $r_{j}^{(\pm )}(k)$ and hence the quantities $%
\Delta _{j}(k)$ are smooth random functions of $j$, i.e. of the distance
along the sample. As the length $N$ of a sample is large enough, $N\gg N_{%
\mathrm{loc}}$, the number of regions where $\mathrm{Re}\,\Delta _{j}(k)>0$
is approximately equal to the number of regions where $\mathrm{Re}\,\Delta
_{j}(k)<0$. The characteristic size of these areas is the localization
length. Therefore the expected \textit{number of cavities}, $N_{\mathrm{cav}}$,
resonant at a given wave number $k$ is: 
\begin{equation}
N_{\mathrm{cav}}\simeq \frac{\widetilde{L}}{2l_{\mathrm{loc}}}.  \label{eq13}
\end{equation}

In order to estimate the \textit{number of resonances}, $N_{\mathrm{res}}$,  in a given
interval $\Delta k$ of the wave numbers (in a given frequency interval $%
\Delta \omega =c\Delta k$), it is necessary to estimate the cavity
``width'', $\delta k_{\mathrm{res}}$, in the $k$-domain. To do this, we note that
the variation $\delta k$ of the wave number leads to a variation $\delta \varphi
_{j}$ of $\varphi _{j}\equiv \arg \Delta _{j}(k)$: 
\begin{eqnarray}
\delta \varphi _{j}=\delta k\frac{d\varphi _{j}}{dk}=\mathrm{Im}\left\{\frac{%
1}{\Delta _{j}(k)}\frac{d\Delta _{j}(k)}{dk}\right\}=\nonumber\\
\delta \kappa \left[ 
\mathrm{Im} \frac{1}{r_{j}^{(+)}(k)}\frac{dr_{j}^{(+)}(k)}{dk}+\mathrm{Im}%
\frac{1}{r_{j}^{(-)}(k)}\frac{dr_{j}^{(-)}(k)}{dk}\right].  \label{eq14}
\end{eqnarray}
The resonant wave number $k_{r}$ is defined by the condition (\ref{DeltaIm}), therefore the second resonance appears in a small vicinity of the same
layer $j$ when the variation of $\arg \Delta _{j}(k)$ approaches $2\pi $: $%
\delta \varphi _{j}\simeq 2\pi $. It is easy to see that the largest
contribution to $\delta \varphi _{j}$ comes from the layers that are the
most distant from the $j$-th layer: 
\begin{eqnarray}
\mathrm{Im}{\frac{d}{dk}}\left[ \log r_{j}^{(\pm )}(k)\right] \sim |%
\widetilde{z}_{j}-\widetilde{z}_{j\pm {N}_{\mathrm{loc}}}| \sim \nonumber\\
{N}_{\mathrm{loc}}\widetilde{d}_{0}=l_{\mathrm{loc}}.  \label{eq15}
\end{eqnarray}
Thus, $\delta \varphi _{j}(k)\sim 2\,\delta k\,l_{\mathrm{loc}}$ and the
characteristic interval $\delta k_{\mathrm{res}}$ between resonant wave
numbers localized around an arbitrary point $\widetilde{z}_{j}$\ can be
estimated as 
\begin{equation}
\delta k_{\mathrm{res}}\simeq \frac{\pi }{l_{\mathrm{loc}}}.  \label{eq16}
\end{equation}
In \cite{JOSA} this result has been obtained for modes located around the
center of the sample.

Equations~(\ref{eq13}) and (\ref{eq16}) allow estimating the number of
resonances $N_{\mathrm{res}}$ in the given frequency interval $\Delta \omega
=c\Delta k$ in the sample of the length ${N}$: 
\begin{equation}
N_{\mathrm{res}}=N_{\mathrm{cav}}\frac{\Delta k}{\delta k_{\mathrm{res}}}%
\simeq {N}\Delta k \widetilde{d_{0}}/2\pi.  \label{eq17}
\end{equation}

The number of cavities, $N_{\mathrm{cav}}$, and the spacing between the resonances, $\delta k_{\mathrm{res}}$, are inversely proportional \cite{JOSA} to $l_{\mathrm{loc}}$, i.e., it increases when the strength of the scattering increases.  Therefore, it follows from Eq.~(\ref{eq17}) that $N_{\mathrm{res}}$ does not depend on the
localization length. In other words, the number of resonances (and therefore
the number of peaks in the transmission spectrum and the number of regions
with enhanced intensity) in a given frequency interval are proportional to
the size of the random system and are independent of the strength of
disorder.
 
An important point is that Eq.~(\ref{eq17}) gives the \textit{total} number
of disorder-induced resonances existing along the whole length
of a random system in a given range $\Delta k$. When random lasing is concerned, each sample is an
active medium, and a new parameter -- the specific gain rate $g$ -- should be
involved. In evaluating the number of \textit{lasing} modes, this parameter has to be compared with the lasing
threshold $g_{c}^{(j)}$ of each
cavity, which is different for different ones. As it has been shown in \cite{JOSA} and mentioned in the Introduction above,  each disorder-induced
resonator occupying an area  $z_{j-N_{s}}\leq z\leq z_{j+N_{s}}$ is built of
strongly reflecting (as a result of Anderson localization) effective
barriers that confine an almost transparent (for the given resonant wave
number $k$) region. Reflection coefficients of the left and right  (from
the center) parts of this structure are large, which
means that the normalized amplitudes $f_{\mp 2k}^{(\pm )}$ 
of the $\pm 2k$-harmonics of the distributions $F_{j}^{(\pm )}$ [Eqs.~(\ref{Distribution_1}), (\ref{Distribution_2})] of the scatterers in these parts of $j$-th cavity are large, $\left|f_{\mp 2k}^{(\pm )}\right|\sim 1$. However, the value $f_{2k}$ of the amplitude of the $2k$-harmonics of the
distribution $F_{j}(z)=F_{j}^{(+)}(z)-F_{j}^{(-)}(z)$ of the scatterers along
the\textit{\ whole} cavity, can be small, or even equal to zero. Indeed,
since 
\[
f_{2k}={f_{-2k}^{(+)}}^{\ast }+f_{2k}^{(-)}={r^{(+)}}^{\ast }-r^{(-)},
\]
\[
\left\vert f_{2k}\right\vert ^{2}=\left\vert r^{(+)}\right\vert
^{2}+\left\vert r^{(-)}\right\vert ^{2}-2\mathrm{Re}\,\Delta (k),
\]
and
\[
\left|r^{(+)}\right|\sim \left|r^{(-)}\right|\sim 1,
\]%
the value of  $\left|f_{2k}\right|$ is small when ${\rm Re}\,\Delta (k)$
is close to unity. For given values of $\left\vert r^{(+)}\right\vert $
and $\left\vert r^{(-)}\right\vert$, the amplitude $\left\vert
f_{2k}\right\vert $ is smaller the greater is ${\rm Re}\,\Delta (k)$. In the
extreme case, the amplitude $\left\vert f_{2k}\right\vert ^{2}$ turns to  zero, i.e.,
the $2k$-harmonics is completely suppressed. This
happens when the amplitudes $f_{\mp
2k}^{(\pm )}$ have equal absolute values but opposite signs, i.e. are
shifted in phase by $\pi$. Thus, the (positive) value of ${\rm Re}\,\Delta (k)$
can be treated as a measure of the amplitude of the $2k$-harmonics of the
cavity and of the phase shift between the $2k$-harmonics of its left and
right parts. 

In a real random sample, the $2k$-harmonics of the spatial distributions $F_j^{(\pm)}(z)$ of scatterers are not equal to zero in any effective cavity and for any $k$ (any frequency $\omega$), including the resonant ones. These harmonics provide distributed feedback, as it occurs,
for instance, in conventional (not random) semiconductor distributed
feedback (DFB) laser (see, e.g., \cite{Okai}). As in a conventional DFB
laser with regular Bragg grating (regular periodic spatial distribution of
the scatterers), in a disorder-induced cavity a lasing mode is excited when
the gain rate $g$ of the medium exceeds the value $g_{c}^{(j)}$ of the
threshold of the cavity. The difference between these two cases is that the whole
periodic grating is characterized by a single value of  $g_{c}$,
while different parts of a random configuration have their own different
thresholds $g_{c}^{(j)}$, whose values  are minimal into the cavities. In
just the same way as the phase shift between two identical halves of the
regular grating causes to decrease the threshold in conventional DFB lasers \cite{Okai}, the phase difference between complex amplitudes ${f_{2k}^{(-)}}$ and 
${f_{-2k}^{(+)}}$ reduces the lasing threshold of the $2k$-resonant random
cavity. From the considerations presented in the previous paragraph it
follows that the information about $g_{c}^{(j)}$ is also contained in the
quantities $\Delta_{j}$: the greater  ${\rm Re}\,\Delta _{j}(k)$ is for a given
cavity, the smaller is its threshold $g_{c}^{(j)}$. Thus, only those
cavities  whose thresholds $g_{c}^{(j)}$ are less than the medium gain $g$
(i.e. the values ${\rm Re}\,\Delta _{j}(k)$ exceed the critical value $\Delta _{c}$), contribute to the formation of the lasing spectrum. Therefore the number
of lasing modes (number of lines in the lasing spectrum) is definitely
smaller than the total number of cavities $N_{\mathrm{res}}$.

To conclude this section we note that approach presented above could  be used in studying various wave systems, for example,  THz tunable vortex photonic crystals. Indeed, many calculations on this problem have already been performed  \cite{Franco_1}, including the role of disorder, but this is beyond the scope of this paper.

\section{Numerical simulations \label{Simulation}}

\begin{figure}[htb]
\centering \scalebox{0.4}{\includegraphics{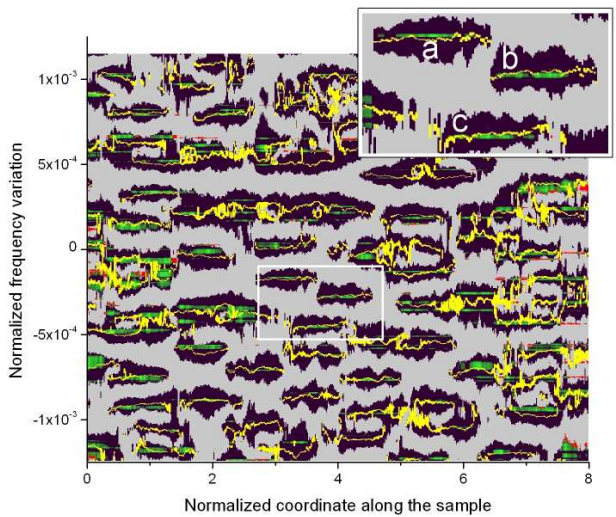}}
\caption{(color online) Mapping of $\Delta_j(k)$ and the normalized resonant
intensities $I_{s}(j,k)$ on the coordinate-frequency plane. The units in
the $x$-axis are normalized to the localization length. In the black areas $\mathrm{Re}\Delta_j(k)>0$, gray color corresponds to $\mathrm{Re}\Delta_j (k)>0$. Approximately half of the full length of any line $k=\omega/c=\mathrm{constant}$ is occupied by black areas where the cavities are located. Yellow color marks the regions where $|\arg \Delta_j(k)-2\protect\pi|<0.05$.
The regions with strong concentration of the wave field, where $I_{s}(j,k)>1/2$, are marked in green. As can be seen, practically all of them are located in the cavities (black areas) as predicted here. Inset: enlarged view of the selected area with three [(a), (b), and (c)] resonances.
}
\label{L-K_plot}
\end{figure}

To verify the results of the previous section, we have calculated the
function $\Delta _{j}(k)$ numerically for different random samples, and have
found the distribution of the areas with $\mathrm{Re}\,\Delta _{j}(k)>0$ and 
$\mathrm{Re}\,\Delta _{j}(k)<0$ along each sample for different wave
numbers. These distributions have been mapped on the coordinate --
wave-number plane in Fig.~\ref{L-K_plot}.

\begin{figure}[tbh]
\centering \scalebox{0.7}{\includegraphics{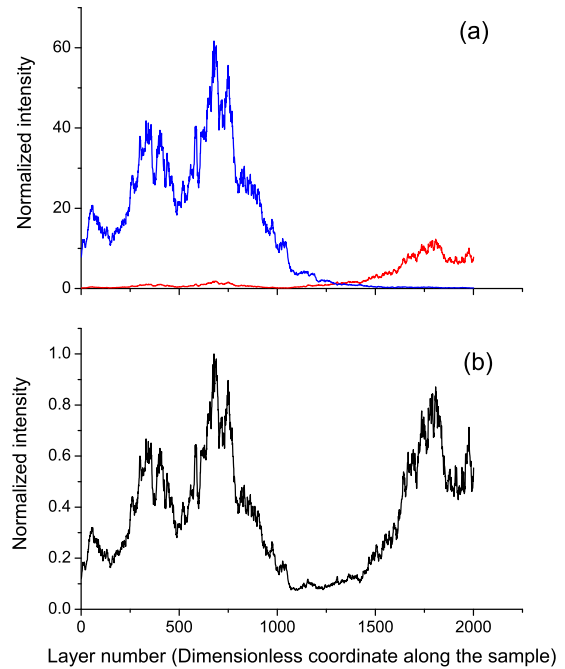}} 
\caption{(color online) (a) Spatial distributions of the intensity 
generated by a wave incident from the left (thin blue) and right (thick red) sides of
the sample; (b) intensity normalized in accordance with Eq.~(\protect\ref{normalized}). The incident wave generates a resonance (peak of the intensity) only in the cavity closest to the edge it is coming from. The distribution of the normalized intensity is independent of the direction of incidence and clearly reveals all cavities resonant at a given frequency.}
\label{Fig_2}
\end{figure}

Then, the eigenvalues, $k_{\mathrm{res}}$, have been determined from Eq.~(%
\ref{Im delta}), and spatial distributions of the intensity created inside
the samples by the incident waves with the corresponding resonant
frequencies ($\omega _{\mathrm{res}}=ck_{\mathrm{res}}$) have been found and
compared with the map. To facilitate this comparison, we take into account
that the intensity pumped by the incident wave into a cavity depends on its
location inside the sample: it decreases exponentially as the distance of
the cavity from the input end increases \cite{JOSA, Payne}. As an example,
the intensity distributions along the same sample illuminated by the same
wave, either from the left $I_{l}(j)$ (blue curve), or from the right, $%
I_{l}(j)$, (red curve) ends are shown in Fig.~\ref{Fig_2}a.

In both curves shown in Fig.~\ref{Fig_2}, the intensity in the closest to
the input cavity is far above the intensity inside the more distant one.
Therefore we have introduced the normalized quantity 
\begin{equation}
I_s(j)={\frac{\left[{I_{l}(j)\over\max \{I_{l}(j)\}}+{I_{r}(j)\over\max \{I_{r}(j)\}}
\right] }{\max \left\{ \left[{I_{l}(j)\over\max \{I_{l}(j)\}}+{I_{r}(j)\over\max
\{I_{r}(j)\}}\right] \right\} }},  \label{normalized}
\end{equation}%
which is symmetric with respect to the direction of incidence and reveals
equally well all resonators, independently on their positions Fig.~\ref%
{Fig_2}b.

In Fig.~\ref{L-K_plot}, regions where $\mathrm{Re}\,\Delta _{j}(k)>0$ and $%
\mathrm{Re}\,\Delta _{j}(k)<0$ are marked by black and gray, respectively.
The normalized intensity $I_{s}(j,k)$ is shown by green. Yellow color marks
the regions where $|\arg \Delta_j(k)-2\pi|<0.05$. One can see that any
horizontal cross-section $k=\mathrm{constant}$ contains approximately the
same number of black and blue regions, and practically all resonances are
located in ``black'' areas associated with cavities, as it was predicted.

\begin{figure}[tbh]
\centering \scalebox{0.8}{\includegraphics{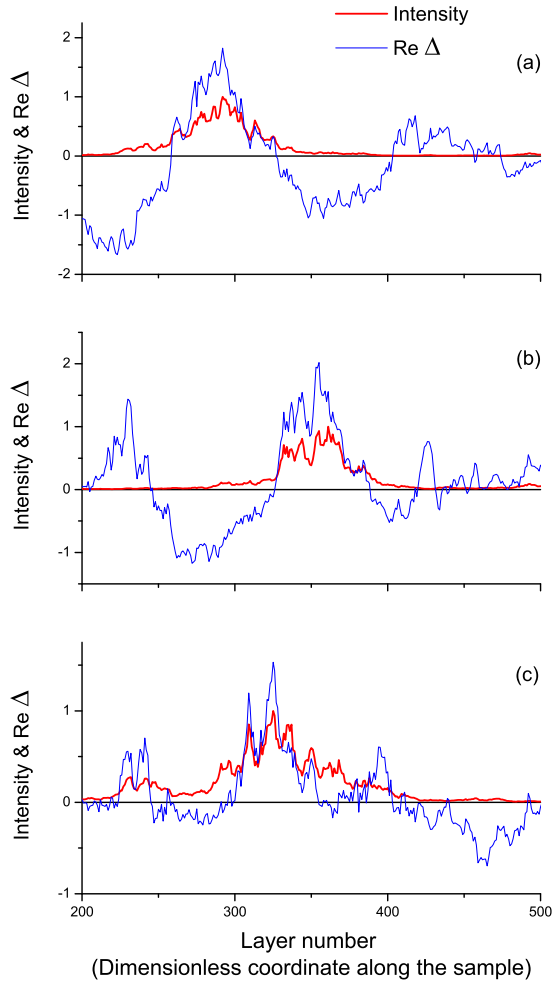}}
\caption{(color online) Spatial distributions of $\mathrm{Re}\,\Delta _{j}$ (thin blue) and of the normalized intensity (thick red) for three resonant wave numbers marked by (a), (b), and (c) in the inset in Fig.~\ref{L-K_plot}. One can see that each resonance is localized in the area where $\mathrm{Re}\,\Delta _{j}$ is positive and takes maximal value.}
\label{Fig.Distribution}
\end{figure}

Examples of the spatial distributions of $\mathrm{Re}\,\Delta _{j}$(blue
curves) for three resonant frequencies marked by (a), (b), and (c) in Fig.~\ref{L-K_plot} are presented in Fig.~\ref{Fig.Distribution}, along with the
corresponding distributions of the intensity (red curves). Shown in Fig.~\ref%
{Correlation} cross-correlation functions, $C(l)=\sum_{j}\mathrm{Re}\,\Delta
_{j}I_{j-l}$, of those two types of curves reveal strong correlation between
the normalized intensity and $\mathrm{Re}\,\Delta$. One can see that the
cavities are well detected by the $\mathrm{Re}\,\Delta >0$ criterion.

\begin{figure}[tbh]
\centering \scalebox{0.8}{\includegraphics{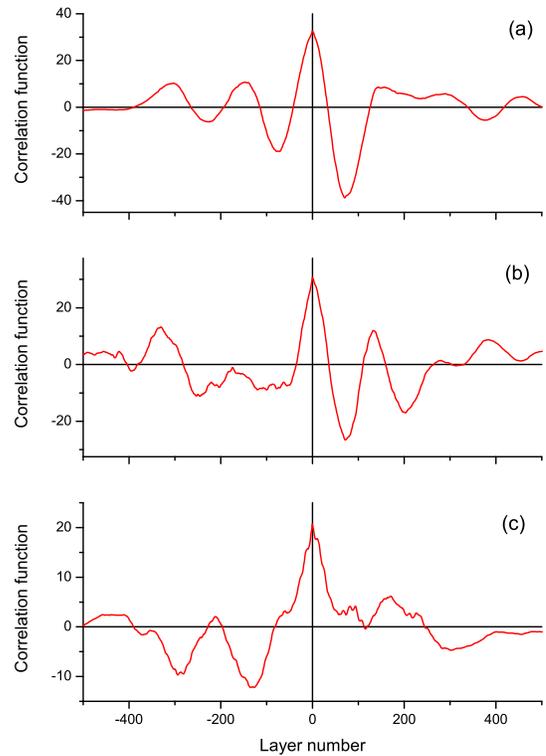}}
\caption{(color online) Cross-correlation functions $C(l)=\sum_j 
\mathrm{Re}\,\Delta _{j} I_{j-l}$ of the normalized intensity and $\mathrm{Re}\,\Delta$ for three resonant wave numbers marked by (a), (b), and (c) in the inset in Fig.~\ref{L-K_plot}. The curves reveal strong correlation between the normalized intensity and $\mathrm{Re}\,\Delta$.}
\label{Correlation}
\end{figure}

\begin{figure}[tbh]
\centering \scalebox{0.8}{\includegraphics{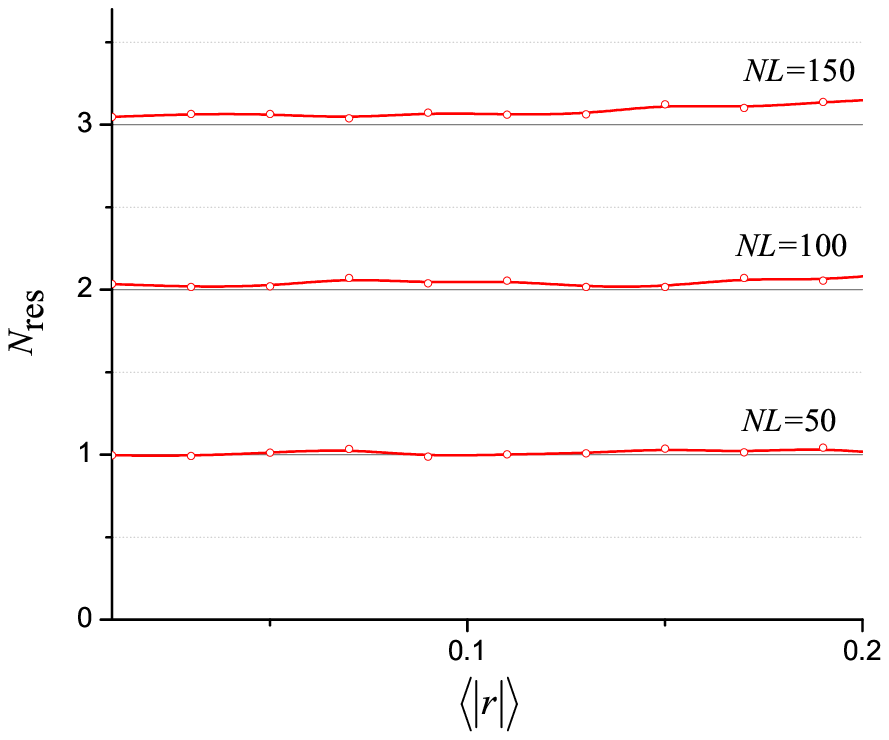}}
\caption{(color online) Number of resonances $N_{\mathrm{res}}$ as a
function of the averaged local reflection coefficient (strength of disorder) 
$\langle|r|\rangle$.  It is easy to see that, as predicted by the theory, the number of resonances in a given frequency interval grows linearly with the size of the sample and is independent of the strength of the disorder. Parameters of the numerical simulations: $k\widetilde{d}_0=10.0$, $\Delta k\widetilde{d}_0=0.1$. Each point is obtained by averaging over $10^3$ random samples.}
\label{Fig.numer_simulation_Nres}
\end{figure}

Figure~\ref{Fig.numer_simulation_Nres} demonstrates that, in accordance with
Eq.~(\ref{eq17}), the number of resonances in a given frequency interval is
independent of the strength of disorder (of the values of the local
reflection coefficients $r_{j}$) and of the variance of the fluctuations of
the thicknesses $d_{j}$, and is completely determined by the size of the
sample.

\begin{figure}[tbh]
\centering \scalebox{0.4}{\includegraphics{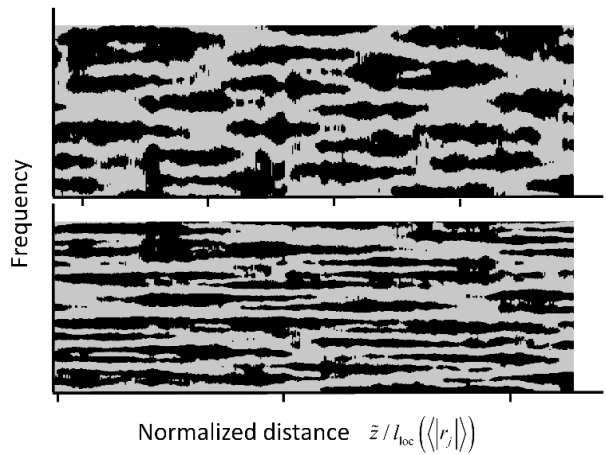}}
\caption{Distribution of the resonant cavities in the coordinate-frequency plane for two geometrically identical samples with different strengths of the scattering and thus with different localization lengths. Top: $\langle|r_j|\rangle=0.2$; bottom: $\langle|r_j|\rangle=0.15$. The distances between the black tick marks under each picture are equal to the corresponding localization lengths.}
\label{Fig_new}
\end{figure}

This rather counter-intuitive result is also supported by the numerical simulations presented in Fig.~\ref{Fig_new}, where the locations of the resonant cavities in the coordinate-frequency plane are shown (marked in black) for two samples. The samples are geometrically identical, i.e. have the same spatial distribution of the scatterers, and differ only in the amplitudes of their reflection coefficients, so that the localization length at the upper picture is twice larger than in the lower panel.
When comparing both images in Fig.~\ref{Fig_new}, it is easy to see that in passing from one picture to another the sizes of the black areas in the $x$-direction increase while the distances between them along  the $y$-axis decrease in the same proportion, so that, in keeping with Eq.~(\ref{eq17}), $N_{\mathrm {res}}$ remains the same, meaning that there is no dependence of the number of resonances on the localization length.

\section{Experiments with active random samples: Number of lasing modes 
\label{Experiment}}

The experimental detection of disorder-induced cavities and resonances is a
challenging task in optics. Although the intensity distribution cannot be
measured directly, the resonances, in principle, can be revealed by
transmission experiments at different frequencies. However, although
resonant transmission in lossless systems is essentially higher than at
typical frequencies, it is much stronger affected by absorption, which is
proportional to the exponentially large intensity inside resonators. In
microwave experiments \cite{Genack we 1, Genack we 2}, where the absorption
length was much larger than the total length of the system (single-mode
waveguide), the transmission was bellow the noise level even at centrally
located resonances. This fact was of little concern in that case because the
microwave probe could be inserted in different points inside the waveguide.
In fiber optic systems, however, this is not possible. With long optical
fibers, to make the transmission measurable it might be
necessary to compensate for the absorption by, for example, introducing
amplification. When the amplification is sufficiently large, all resonances
manifest themselves as sharp lines in the transmission spectrum.

Experimental studies of resonances in disordered optical fibers with active
elements is also of interest for a better understanding of the physics of
RL, because in amplifying media all resonant modes are potential candidate
for lasing. The ability to monitor and to tailor their parameters,
especially the total number, eigen frequencies, and locations is critical to
fulfill this potential.

Random one-dimensional cavities can be created in optical fibers by the
introduction of randomly-positioned reflectors, in the form of Bragg
gratings \cite{2005_Shapira}. To create such structures, we have used
commercial Er/Ge co-doped fibers (INO Quebec, QC, Canada) that are
single-mode at 1535 nm. Erbium is the active element and germanium doping
provides a kind of photosensitivity that can be used to change, locally, the
refractive index. The Bragg gratings were fabricated by exposing the fiber
to UV light (244 nm) from a frequency doubled argon ion laser through a
periodic mask whose spatial period is 1059.8nm. Each one of the fabricated
gratings had a length of approximately $5\,$mm. The random distances between
the gratings were statistically independent and uniformly distributed in the
interval $d_{0}\pm 0.8\,$mm, where the mean distance between gratings $d_{0}$
was approximately $5\,$mm.

The fabricated Bragg grating have a narrow reflection spectrum (its full
width of half maximum is about 0.17 nm) centered at $1535.3 \,$nm, with a
maximum reflectivity of about 0.07-0.08. Under these conditions, the estimated localization length is found to be about 5-6 gratings. We notice  that variations in the mask alignment,
recording exposure, and fiber tension during the writing process caused
small variations in the central wavelength of the gratings and on the
sharpness of their reflection spectra. As a result, the half width of the
reflection spectrum of an array of 31 gratings is about $0.27\,$nm.

\begin{figure}[tbh]
\centering \scalebox{0.4}{\includegraphics{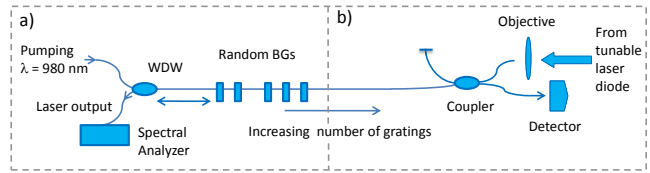}}
\caption{(color online) Schematic diagram of the experimental setup. In (a) we illustrate the configuration of the laser. The Wavelength-Division Multiplexing (WDM) device permits the separation of the pump ($980 \,$nm) and lasing radiation ($1550 \,$nm). In (b) we show the arrangement used to measure the reflection spectrum of the grating array during the fabrication process.}
\label{Fig.setup}
\end{figure}

The optical arrangement employed to fabricate the gratings and to measure the
reflection spectra of the arrays is illustrated in Fig.~\ref{Fig.setup}.
Laser action was obtained by end-pumping the system with 980 nm radiation
from a semiconductor laser. A wavelength-division multiplexing (WDM) was
used to separate the pumping wavelength from the radiation emitted by the
laser (see Fig \ref{Fig.setup}). 
Measurements of the gratings transmission/reflection coefficients with a spectral resolution of $0.001\,$nm were carried out in the spectral range $1520$--$1580 \,$nm, using a tunable semiconductor laser (New Focus Velocity-6300) with a coherence length of a few meters.
As illustrated in the figure, new gratings
were fabricated in the sequence, beginning from the pumping end of the fiber.

\begin{figure}[tbh]
\centering \scalebox{0.65}{\includegraphics{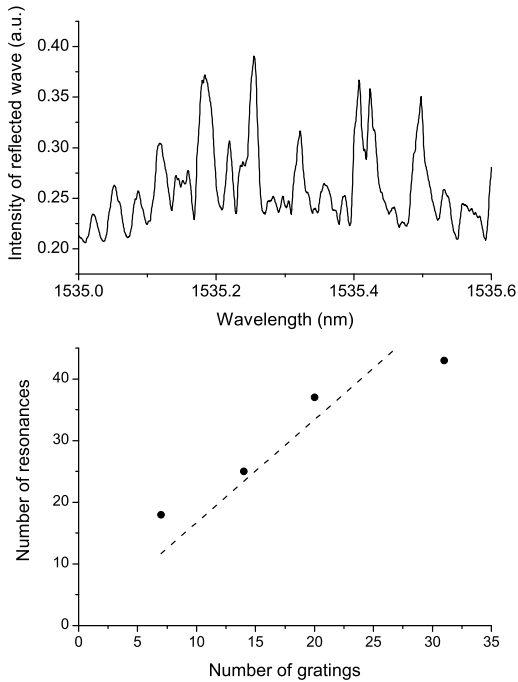}}
\caption{Top: Experimental reflection spectrum from an array of fourteen gratings. 
Bottom: Experimental (black points) and theoretical (dotted line)
dependence of the number of resonances, $N_{\mathrm{res}}$, on the number of
gratings $N_{\mathrm{g}}$. The theoretical dependence is calculated using Eq.~\ref{eq17} and the parameters of the experiments.}
\label{NTransExp}
\end{figure}

To explore the dependence of the total number of resonances, $N_{\mathrm{res}%
}$, on the size of the system, we measured the frequency spectrum of the
reflection coefficient at samples with different numbers of reflectors, and
made use of the fact that each resonance manifested itself as a sharp drop
of the reflectivity. Black points in Fig.~\ref{NTransExp} represent the
total number of the resonances detected in the arrays of different numbers
of gratings $N_{g}$ (different lengths of the samples, $L=N_{g}d_{0}$, $%
d_{0}=5\,$mm) in the wavelength range ($1534.8$--$1535.6\,$nm). The
theoretical prediction Eq.~(\ref{eq17}), $N_{\mathrm{res}}=1.67N_{g}$
(dotted line), is in an excellent agreement with the experimental data.

\begin{figure}[tbh]
\centering \scalebox{0.4}{\includegraphics{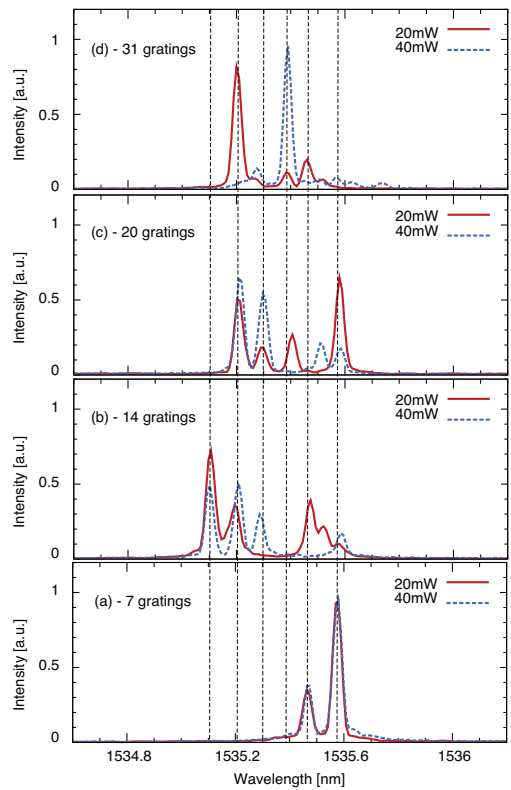}}
\caption{(color online) Emission spectra of the samples with 7, 14, 20, and
31 gratings for 20 mW (green) and 40 mW pumping powers. Both values are well
above the lasing threshold. The vertical lines mark the spectral positions
of the emission lines. }
\label{laser_emission}
\end{figure}

In Fig.~\ref{laser_emission}, we present the results of the measurements of
the emission spectra of the RL fiber containing 7, 14, 20, and 31 randomly
distributed Bragg gratings (RDBG), for two values of the pumping power: 20
mW denoted by the continuous line curves, and $40\,$mW, denoted by the
dashed line curves. With these pump levels, the systems were above threshold
in all cases. For the measurements we used a spectrum analyzer with the
resolution of $0.01\,$ nm.

One can see from Fig.~\ref{laser_emission} that in the considered frequency
range, the emission spectrum contains several peaks that reveal the presence
of resonant modes. As the number of gratings grows so does the number of
resonances: there are two peaks for the RL with seven gratings and seven peaks
for the sample with 20 gratings. For the RL with seven gratings, the two peaks
maintain their positions and relative intensities as the pump power
increases. For systems with a higher number of gratings, the competition
between modes produces temporal fluctuations in the relative strengths of
the emission lines; these fluctuations also depend on the pump power. While
the relative intensity of the peaks can change with the pump power, their
spectral positions (indicated by the vertical lines in the figure) remain
fixed.

Another interesting feature that can be observed in Fig.~\ref{laser_emission}
is that, once an emission line appears in a system with a low number of
gratings, it is likely to reappear in a system with a higher number of
gratings. One can see, for example, that the emission lines observed with
the system with seven gratings are also present in the systems with more
gratings.

\begin{figure}[tbh]
\centering \scalebox{0.55}{\includegraphics{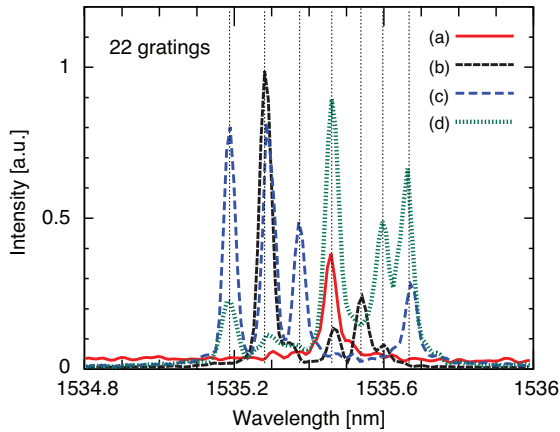}}
\caption{(color online) Emission spectra of the sample with 22 gratings
generated by the fixed pumping power (40 mW) at different moments of time.}
\label{Fig.time_variation}
\end{figure}

The curves shown in Fig \ref{Fig.time_variation} represent spectra of the
radiation emitted by a random laser with 22 gratings, measured in 1 second
intervals. Even at constant pumping power, the number of well-defined lasing
modes and their emission intensity fluctuate in time. The emission
frequencies, however, remain fixed; one can see that they always coincide
with one of the vertical lines of the figure. The observed fluctuations of
the intensity of the spectral lines could not be caused by relatively small ($\sim$5\%) fluctuations of the intensity of the pumping laser, and apparently were associated with non-linear effects. Indeed, despite the relatively low power of the emission, the field inside the high-$Q$ cavities can be  strong enough (due to resonance) to generate a Kerr-type nonlinearity. 
These effects are of importance, for example, in distributed Bragg reflector fiber lasers that are longer than $20\,$cm \cite{Lian, Liu}

\begin{figure}[th]
\centering \scalebox{0.8}{\includegraphics{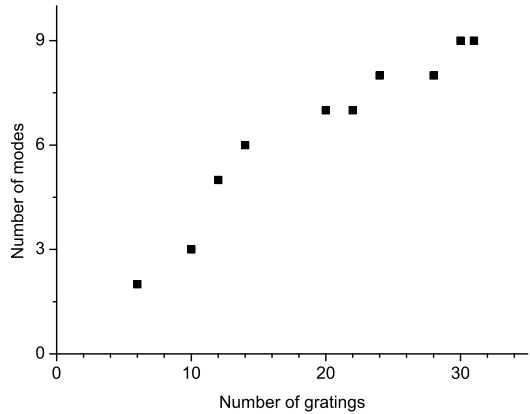}}
\caption{Number of resonances measured in the wavelength
range $\protect\lambda \pm \Delta \protect\lambda =1.5\cdot 10^{-4}\pm
0.85\cdot 10^{-8}$cm (black points) as a function of the number of Bragg
reflectors. Since the total length of the sample is proportional to the
number of gratings, the figure actually presents the dependence on size of
the random system.}
\label{Fig.numer_exp_theory}
\end{figure}

The black points in Fig.\ \ref{Fig.numer_exp_theory} denote the number of
lasing modes measured in the wavelength range $\lambda \pm \Delta \lambda
=1535.3\pm 0.3\,$nm in samples with different numbers of gratings $N_{g}$
(and different lengths of the system; $L=N_{g}d_{0}$, $d_{0}=5\,$mm). These
numbers were obtained by adding all the emission lines present in the
emission spectra over an extended period of time. The amount of the lasing
modes growths linearly with the length of the system, although, in
accordance with the theoretical reasoning above, it is always less than the
total number of the effective resonant cavities in the sample (Fig.~\ref%
{NTransExp}).

\section{Summary}

To conclude, eigen-modes (resonances) of a randomly layered long ($L\gg l_{%
\mathrm{loc}}$) sample are localized in disorder-induced effective cavities
of the size of the order of the localization length that are randomly
distributed along the sample. An algorithm for finding this distribution in
an individual configuration with arbitrary (random) parameters was
developed based on the calculation of the function $\Delta_{j}(k)$ in Eq.~(%
\ref{delta}). It was shown that a cavity of an effective size of a few
localization lengths, in which a mode with $k=k_{\mathrm{res}}$ can be
localized, exists around $j$-th layer when $\mathrm{Im}\,\Delta _{j}(k_{%
\mathrm{res}})=0$ and $\mathrm{Re}\,\Delta _{j}(k_{\mathrm{res}})>0.$ In the
case of uncorrelated disorder and weak scattering, the spacing between
eigen-levels and the number of cavities $N_{\mathrm{res}}$ in a given
frequency interval does not depend on the strength of disorder, and are
uniquely determined by the size of the sample. The frequency of each
resonance is independent of the coordinate of the effective cavity, in which
it is located. The number of lasing modes depends also on
the ratio between the threshold values $g_{c}^{(j)}$ of
the individual cavities and the gain $g$ of the medium, and is less than $N_{\mathrm{res}}$. The theoretical predictions and numerical results are in
reasonable agreement with the experimental data obtained by measuring the
emission spectra of the random laser based on the single-mode fiber with
randomly distributed Bragg gratings.

\begin{acknowledgments}

FN is partially supported by the ARO,
JSPS-RFBR contract No. 12-02-92100,
Grant-in-Aid for Scientific Research (S),
MEXT Kakenhi on Quantum Cybernetics,
and the JSPS via its FIRST program. 
VF is partially supported by the Israeli Science Foundation (Grant No. 894/10).
We thank Dr. Sahin Ozdemir for his careful reading of the manuscript and useful comments.

\end{acknowledgments}

.

\end{document}